\documentclass[aps,prl,twocolumn,showpacs]{revtex4}

\usepackage{graphicx}
\usepackage{dcolumn}
\usepackage{amsmath}
\usepackage{amssymb}
\usepackage{wasysym}
\usepackage{color}
\begin{document}

\title{Direct observation of multiple spin zeroes in the underdoped high temperature superconductor YBa$_2$Cu$_3$O$_{6+x}$}

\author{Suchitra~E.~Sebastian$^1$}
\email{suchitra@phy.cam.ac.uk}
\author{N.~Harrison$^2$}
\email{nharrison@lanl.gov}
\author{M.~M.~Altarawneh$^2$}
\author{F.~F.~Balakirev$^2$}
\author{Ruixing Liang$^{3,4}$}
\author{D.~A.~Bonn$^{3,4}$}
\author{W.~N.~Hardy$^{3,4}$}
\author{G.~G.~Lonzarich$^1$}

\affiliation{
$^1$Cavendish Laboratory, Cambridge University, JJ Thomson Avenue, Cambridge CB3~OHE, U.K\\
$^2$National High Magnetic Field Laboratory, LANL, Los Alamos, NM 87545\\
$^3$Department of Physics and Astronomy, University of British Columbia, Vancouver V6T 1Z4, Canada\\
$^4$Canadian Institute for Advanced Research, Toronto M5G 1Z8, Canada
}
\date{\today}

\begin{abstract}
We report the direct observation of multiple `spin zeroes' in angle-dependent magnetic quantum oscillations measured up to 85T in YBa$_2$Cu$_3$O$_{6+x}$, at which the amplitude falls to a deep minimum accompanied by a phase inversion of the measured quantum oscillations, enabling the product of the effective mass and effective {\it g}-factor $m^\ast g^\ast$ to be tightly constrained. We find an evolution of the location of the spin zeros with applied magnetic field, and suggest that this effect and the absence of a spin zero at low angles can be produced by more than one Fermi surface component, and an effective {\it g}-factor with a subtle anisotropy between in-plane and out-of-plane crystalline directions.
\end{abstract}
\maketitle

Magnetic quantum oscillations measurements made in the underdoped cuprates YBa$_2$Cu$_3$O$_{6+x}$~\cite{doiron1,leboeuf1,jaudet1,sebastian1,audouard1,sebastian2,sebastian3,singleton1,sebastian4,sebastian5,ramshaw1,riggs1} and YBa$_2$Cu$_4$O$_8$\cite{yelland1,bangura1} reveal pockets of carriers greatly reduced in size compared to the large paramagnetic Fermi surface observed in the overdoped regime \cite{vignolle1}. Translational symmetry breaking due to a conventional order parameter such as a spin-density-wave or charge-density-wave~\cite{millis1}, or an unconventional order parameter such as d-density-wave order~\cite{chakravarty1}, among others has been suggested to underlie the small Fermi surface pocket size. Yet in the absence of the direct observation of any such long range order, it has proved challenging to distinguish between various potential order parameters on the basis of compatibility with the observed quantum oscillation properties.

\begin{figure}
\centering 
\includegraphics*[width=.46\textwidth]{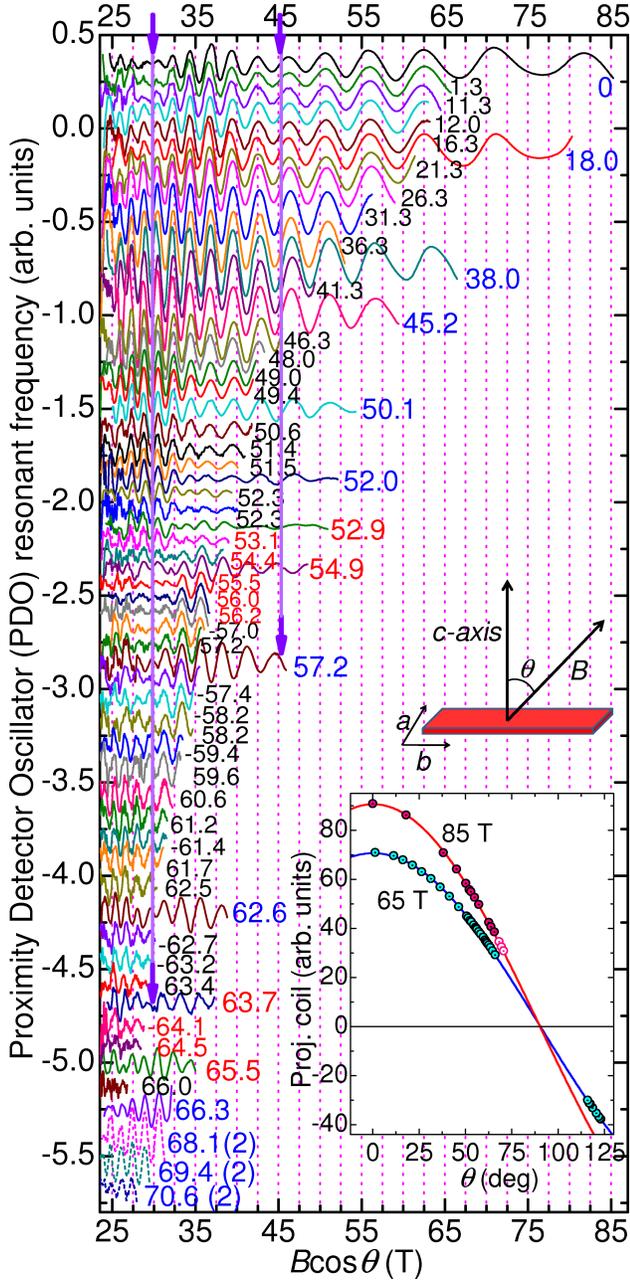}
\caption{Quantum oscillations in the PDO resonance frequency at different angles plotted as a function of $B\cos\theta$. Data have been divided by the damping factor $\exp(-\Gamma/B\cos\theta)$ where $\Gamma=$~170~T for ease of visual comparison. Data collected with 85~T pulses (the highest field used) are labelled by large blue numbers, while data near the two spin zeros are labelled by large red numbers. Two vertical lines at $B\cos\theta\approx 30$ T and  $\approx 45$ T denote the start and end of a beat minimum, which remain approximately the same between 0$^\circ$ and 63.7$^\circ$ except in the vicinity of spin zeroes. The upper inset shows a schematic of the sample configuration ($\theta$ measured with respect to the crystalline $c$-axis). The lower inset shows the $B\cos\theta$ component of the magnetic induction along the c-axis of the sample for 65~T (cyan circles) and 85~T (pink circles), obtained using the projection coil. The angular error is $<0.2^\circ$ for $\theta \leq 66.3^\circ$ and $\approx 0.2^\circ$ for $68.1^\circ \leq \theta \leq 70.6^\circ$ (implying a phase uncertainty at the highest angles $-$ shown by dotted lines).}
\label{data}
\end{figure}

One way to experimentally separate the effects of spin and orbital degrees of freedom associated with the conduction electrons constituting the Fermi surface is to rotate the orientation of the magnetic field in quantum oscillation measurements to detect the presence of signature `spin zeroes' arising from the interference between spin-up and spin-down components~\cite{shoenberg1}. Previous studies on underdoped YBa$_2$Cu$_3$O$_{6+x}$~\cite{sebastian2,ramshaw1}, however, have been restricted to angles $\lessapprox 57^\circ$, yielding different results as to whether one spin zero or none is observed within this angular range. Furthermore, a unique value of renormalised `$g$'-factor $g^\ast=\frac{g_{\textrm s}}{1+F^0_a}$ independent of other assumptions about the Fermi surface topology [where $g_{\textrm s}$ is the `$g$'-factor of the electrons at the Fermi surface and $F^0_a$ is the Landau Fermi liquid coefficient (negative for repulsive interactions) comprising both electron-electron and electron-phonon interactions~\cite{pines1}] can only be obtained unambiguously by the experimental observation of more than a single `spin zero'~\cite{shoenberg1}.

Here we trace quantum oscillations in underdoped YBa$_2$Cu$_3$O$_{6+x}$ in angles up to $\theta\approx 71^\circ$ between the crystalline c-axis and the applied magnetic field ($\mu_0H$) using fields extending to 85~T and detect two well-defined `spin zeroes' in both the amplitude and phase. With this observation we can tightly constrain the value of $g^\ast$ for the Fermi surface section corresponding to the dominant spectral frequency between $\approx$~1.6 and $\approx$~1.9, independent of assumptions about the topology of the Fermi surface. While the value of the unrenormalised `$g$'-factor ($g_{\textrm s} \leq g^\ast$) is contingent on the size of many-body interactions in the strongly correlated cuprates~\cite{pines1}, a strongly suppressed value of $g_{\textrm s}$ (as expected for certain spin-density-wave models, for example~\cite{ramazashvili1, kabanov1}, appears to be ruled out.)

Detwinned single crystals of YBa$_2$Cu$_3$O$_{6.56}$ of dimensions 0.5~x~0.8~x~0.1~mm$^3$ were grown and prepared at the University of British Columbia~\cite{liang1}. Quantum oscillations were measured at the National High Magnetic Field Laboratory (Los Alamos) using the contactless conductivity technique (described elsewhere~\cite{altarawneh1,sebastian3,sebastian4,sebastian5}) with both the proximity detector coil and sample to which it is coupled rotated {\it in situ}.  The sample temperature is maintained close to $T\approx$~1.5~K throughout the experiment by direct immersion in superfluid $^4$He. A worm drive-driven rotator powered by a stepper motor is used for sample rotation, with a secondary angular calibration provided by a pancake projection coil wound in the plane of the sample. The measured component of the magnetic induction $B\cos\theta$ (where $B\approx\mu_0H$) projected along the crystalline $c$-axis shown in Fig.~\ref{data} (inset) yields an uncertainty of $\lessapprox$~0.2$^\circ$ in the sample orientation.

The value of $g^\ast$ is probed in quantum oscillation experiments in quasi-two-dimensional metals by rotating $\theta$, causing the ratio of the Zeeman energy to the cyclotron energy to change~\cite{mckenzie1,shoenberg1}. For simplicity, we consider first the quantum oscillations and orbitally averaged value of $g^\ast$ associated with a single sheet of the Fermi surface with cyclotron effective mass $m^\ast$ (for $\theta=0$). The spin splitting factor~\cite{shoenberg1} $R_{\textrm s}=\cos\big[\frac{\pi}{2} \frac{m^\ast g^{\ast}}{m_{\rm e}\cos\theta}\big]$ in the quantum oscillation amplitude is exactly equal to 0 at certain special angles
\begin{equation}\label{spinzeroes}
\theta_{\textrm sz}=\cos^{-1}\bigg[\frac{g^\ast m^\ast}{(2n+1)m_{\rm e}}\bigg],
\end{equation}
termed `spin zeroes,' where $n$ is an integer. Destructive interference between quantum oscillations from two spin channels gives rise a strong suppression of the amplitude accompanied by change in their phase by $\pi$ on passing through $\theta_{\textrm sz}$.

Figure~\ref{data} shows magnetic quantum oscillations in the resonance frequency shift of the proximity detector oscillator (PDO) circuit plotted versus $B\cos\theta$. An unambiguous first spin zero is seen to occur at $\theta_{\textrm sz}^{\textrm {first}} \approx 53^\circ$ for $B \cos\theta\gtrapprox$~33T in the measured data, identified by a vanishing of amplitude shown in figure 1 accompanied by a phase inversion of quantum oscillations by $\pi$ at lower and higher angles (consistent with the phase inversion observed between 51$^\circ$ and 57$^\circ$ reported in \cite{ramshaw1}). We find a shift in the location of the first spin zero to a higher angle $\theta_{\textrm sz}^{\textrm {first}} \approx 57^\circ$ for  $B \cos \theta \lessapprox$ 33T, thereby reconciling the current observation with the previous differences found between 
experiments performed in low continuous magnetic fields~\cite{sebastian2} and high pulsed magnetic fields~\cite{ramshaw1}. 

\begin{figure}
\centering 
\includegraphics*[width=.46\textwidth]{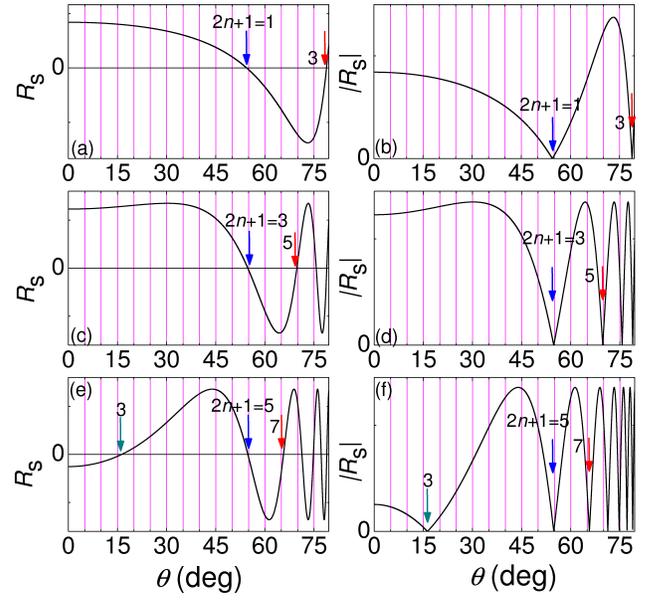}
\caption{$R_{\textrm s}$ and $|R_{\textrm s}|$ as indicated for a single quasi-two-dimensional Fermi surface. Figures (a) $\&$ (b), (c) $\&$ (d), and (e) $\&$ (f) correspond to three very different values of $m^\ast g^\ast \approx 0.6, 1.7,~\&~2.9$ (assuming $g^\ast$ to be isotropic), each of which yield a spin zero at $53^\circ - 57^\circ$ for $2n+1=$~1, 3, or 5 respectively.}
\label{Rs}
\end{figure}

From inspection of Eqn.~\ref{spinzeroes} it can be seen that $g^\ast$ is not uniquely determined by a single spin zero (unless the Fermi surface topology and form of all quantum oscillation damping factors are assumed to be known~\cite{incoherence}), since the conditions for $\theta_{\textrm sz}$ (Eqn.~\ref{spinzeroes}) can be satisfied by multiple values of $(2n+1)$. In the case of underdoped YBa$_2$Cu$_3$O$_{6+x}$, the identification of a spin zero location at $53^\circ-57^\circ$ renders possible values of $2n+1$ as 1, 3, or 5 associated with this spin zero $-$ the corresponding forms of $R_{\textrm s}$ and $|R_{\textrm s}|$ for each value of $n$ are shown in Fig.~2. Each of the potential values of $2n+1$ associated with the spin zero at $53^\circ-57^\circ$ can be distinguished only by the location of the second spin zero, which would uniquely identify a value of ${m^\ast}g^\ast$~(see the form of $R_{\textrm s}$ in Eqn.~\ref{spinzeroes}).

\begin{figure}
\centering 
\includegraphics*[width=.46\textwidth]{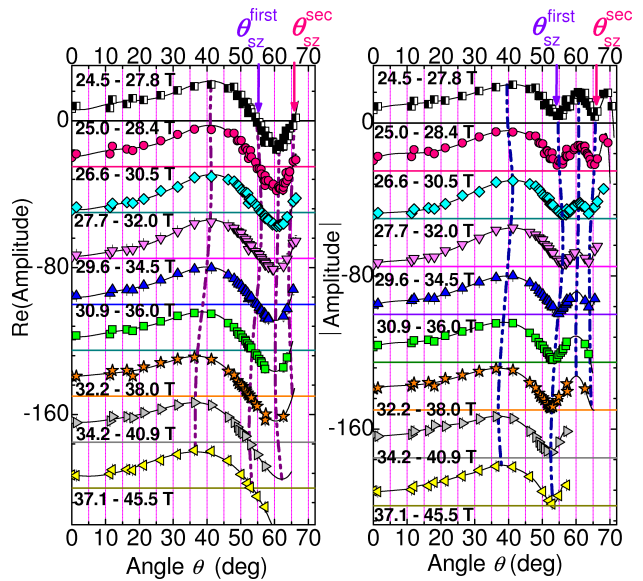}
\caption{{\bf a}, Cross-correlation between the experimental data over various field intervals 26~$<B\cos\theta<$85~T and the sinusoidal function $\cos(2\pi F/B+\phi)$ as described in the text for angles $\theta\leq$~66.3$^\circ$ (for which the angular uncertainty $<~0.2^\circ$). The value of $\phi$ is adjusted to maximize the amplitude (shown in arbitrary units) at each angle $\theta$. {\bf b} Absolute amplitude analysis for the same data extending to $\theta\leq$~71$^\circ$. Dash-dotted lines show the field-evolution of each of the two amplitude maxima and two spin zeros. 
}
\label{correlation}
\end{figure}

Our quantum oscillation measurements at angles up to 71$^\circ$ (Fig.~1) reveal for the first time a second spin zero located at $\theta_{\textrm sz}^{\textrm {sec}} \approx 64^\circ-66^\circ$ evinced by the phase inversion of the oscillations by $\pi$ at lower and higher angles, identifying this spin zero as being associated with $2n+1=7$. To facilitate a precise determination of the phase flip and a comparison of the measured experimental data with the form of $R_{\textrm s}$, we perform a correlation analysis. In Fig.~\ref{correlation} we plot the cross-correlation between the data over different fixed ranges of $B\cos\theta$ (denoted in the figure for each case) for numerous measured angles with a simple sinusoid $\cos(2\pi F/B\cos\theta+\phi)$, where $F$ and $\phi$ are matched to the periodicity and phase of the oscillations over this same limited range at $\theta \approx$~40$^\circ$. The autocorrelation, which does not invoke a simple sinusoidal factor, is expected to lead to similar plots (albeit with more scatter) to those shown in figure~\ref{correlation}. In quasi-two-dimensional metals with weak Fermi surface corrugation~\cite{mckenzie1}, the result of our cross-correlation procedure closely follows the form of $R_{\textrm s}$ (see Fig.~\ref{correlation}). 

As a starting point, we compare the results of the cross-correlation and absolute amplitude analyses obtained from the experimental data with the form of $R_{\textrm s}$ for a single two-dimensional Fermi surface, since a single frequency with minimal warping is known to dominate the measured quantum oscillations~\cite{sebastian5}. From the location of both the absolute amplitude maxima and minima and the zero crossings in phase over the wide range up to 71$^\circ$, close agreement is seen with only a single possibility: $2n+1=5$ corresponding to $\theta_{\textrm sz}^{\textrm {first}}$ at $53^\circ-57^\circ$, and $2n+1=7$ corresponding to $\theta_{\textrm sz}^{\textrm {sec}}$ at $63^\circ-66^\circ$ (Figs.~2e,f). The corresponding value of renormalised {\it g}-factor $g^\ast$ therefore has a unique value determined by the location of the two observed spin zero angles: $1.6 \lessapprox g^\ast \lessapprox 1.9$, and a quasiparticle effective mass $\frac{m^\ast}{m_{\textrm e}}=1.6(1)$~\cite{sebastian5} (where $m_{\rm e}$ is the free electron mass).

There are two ways in which the form of the cross-correlation and absolute amplitude analyses (figure~\ref{correlation}) depart from anticipated form of $R_{\textrm s}$ for a single quasi-two-dimensional Fermi surface. First, the locations of the zero crossings observed in the measured quantum oscillations evolve with magnetic field, with the lower spin zero $\theta_{\textrm sz}^{\textrm {first}}$ falling from $\approx 57^\circ$ to $\approx 53^\circ$ on increasing the magnetic field. The higher spin zero $\theta_{\textrm sz}^{\textrm {sec}}$, on the other hand, rises from $\approx 64^\circ$ to higher angles with increasing magnetic field. Second, for $2n+1=5$ corresponding to the spin zero at $53^\circ-57^\circ$, it can be seen from Figs.~2e and f that a lower angle spin zero is in fact expected to occur at $\approx 16^\circ$ associated with the lower integer $2n+1=3$. In contrast, no spin zero is observed below the first spin zero at $53^\circ-57^\circ$ in the experimentally measured quantum oscillations.

\begin{figure}
\centering 
\includegraphics*[width=.46\textwidth]{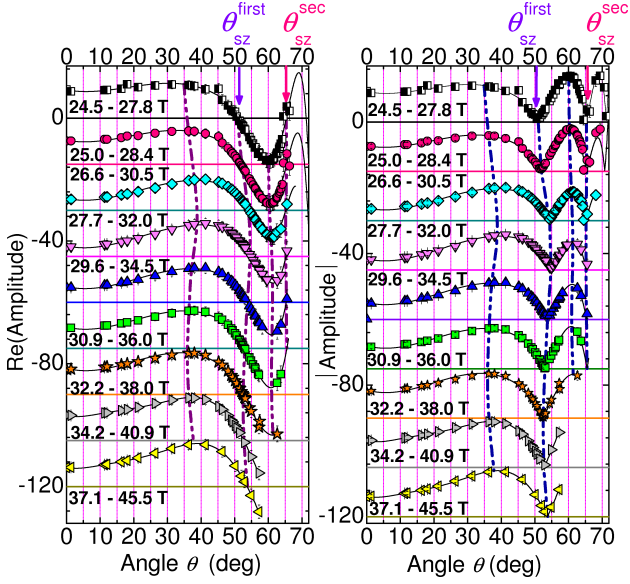}
\caption{Cross-correlation and absolute amplitude analysis on considering two components of the Fermi surface $\alpha$ and $\gamma$ where the quantum oscillation amplitude is given by $a=\sum_ia_iR_{{\rm s},i}R_T\exp\big(\frac{-\Gamma_i}{B\cos\theta}\big){\rm J}_0\big(\frac{2\pi\Delta F_{{\rm w},i}}{B\cos\theta}\big)\cos\big(\frac{2\pi\Delta F_{{\rm s},i}}{B\cos\theta}\big)\cos\big(\frac{2\pi F_i}{B\cos\theta}-\pi\big)$. Here $\Delta F_{{\rm w}_i}=\Delta F_{{\rm w}_i0}{\rm J}(ck_{\rm F}\tan\theta)$ represents a warping of the Fermi surface due to interlayer hopping, and $\Delta F_{{\rm s}_i}$ represents a splitting of the Fermi surface~\cite{lebed1}, $\Gamma_i$ is a damping parameter (such as that caused by scattering~\cite{shoenberg1}) and $R_T=X/\sinh X$ ($X=2\pi m^\ast k_{\rm B}T/e\hbar B$). Shown here are results yielding a reasonable fit to the extended field and angle-dependent data shown in Fig.~\ref{data}, with $F_\alpha=533$T, $\Delta F_{{\rm s},\alpha}=6$T, $g_{z,\alpha}^\ast=1.96$ for the spectrally dominant component, and $F_\gamma=534$T,  $\Delta F_{{\rm s},\gamma}=90$T, $g_{z,\gamma}^\ast=1.42$ for a spectrally weaker component. The best fit is obtained with no appreciable warping ($\Delta F_{{\rm w},\alpha0} \approx \Delta F_{{\rm w},\gamma0} \approx$~0). We consider a simple {\it g}-factor anisotropy of the form $g^\ast=g^\ast_z\sqrt{\cos^2\theta+\frac{1}{\xi}\sin^2\theta}$~\cite{walstedt1}, where $\xi=$~1 for an isotropic {\it g}-factor. Conventional quantum oscillation damping factors~\cite{shoenberg1} included in the above model do not capture an observed $\sim$~30~\%~suppression of the amplitude at large angles $\theta\gtrapprox$~60$^\circ$ in Fig.~\ref{data}~\cite{incoherence}.
}
\label{fits}
\end{figure}

One simple explanation for some of these features is finite $c$-axis dispersion effects such as warping and/or splitting effects~\cite{lebed1}, leading to multiple frequencies~\cite{sebastian3,ramshaw1,audouard1}. We consider the entire set of angular dependent quantum oscillations measured over the broad angular range $0^\circ \leq \theta \leq 71^\circ$ and the extended field range 24~T~$\leq B\leq$~85~T, and attempt to find a set of warping ($\Delta F_{{\rm w},i}$) and/or splitting ($\Delta F_{{\rm s},i}$) parameters to fit the entire data range on considering a minimal model of two Fermi surface components ($i=$~$\alpha$ and $\gamma$) with median frequencies $F_\alpha$ and $F_\gamma$. An interpretation in terms of a predominant splitting rather than warping is suggested by the fixed location of the beat maximum and minimum in field (figure~\ref{data}), as the angle is varied; figure~\ref{fits} shows results on considering such a model.

The contribution from more than a single Fermi surface component can mostly explain the observed field dependence of the zero crossing location, although effects of nonlinearities in the Zeeman splitting cannot be ruled out. The absence of another spin zero occuring at an angle $\theta\approx$~16$^\circ$ (as expected from Figs.~2e and f), however, requires an additional explanation $-$ this absence of a low angle spin zero was previously interpreted as a suppression of the value of $g^\ast$~\cite{sebastian2}. One possibility is an anisotropy in the {\it g}-factor $-$ a simple anisotropy~\cite{walstedt1} of the form $g^\ast=g^\ast_0\sqrt{\cos^2\theta+\frac{1}{\xi}\sin^2\theta}$ anticipated for the CuO$_2$ planes in which $\xi \approx 1.4(1)$ fits the data fairly well.

We conclude that the value of renormalised {\it g}-factor $g^\ast=\frac{g_{\textrm s}}{1+F_{\textrm a}^0}$ of conduction electrons at the Fermi surface section corresponding to the dominant spectral frequency in underdoped YBa$_2$Cu$_3$O$_{\textrm 6+x}$ is tightly constrained to be between 1.6 and 1.9, with $g_{\textrm s} \lessapprox g^\ast$, based on our observation of two spin zeros in quantum oscillations measured over an extended angular range up to 71$^\circ$ in magnetic fields up to 85~T; a strong suppression of the value of $g^\ast$ appears to be ruled out. A potential anisotropy between the in-plane and out-of-plane components of the {\it g}-factor is suggested. The extended set of measured quantum oscillations over a broad angular and field range is consistent with a Fermi surface dominated by splitting effects and negligible warping.

This work is supported by the Royal Society, King's College (Cambridge University), US Department of Energy BES ``Science at 100 T," the National Science Foundation and the State of Florida.


\begin{thebibliography}{99}
\bibitem{incoherence} Other quasi-two-dimensional conductors, like YBa$_2$Cu$_3$O$_{6+x}$, are observed to exhibit some attenuation of the amplitude at large angles~\cite{organics}, possibly due to additional decoherence effects on tilting the orbits slightly out of the plane.
\end{thebibliography}
\end{document}